\documentclass[12pt]{article}

\usepackage{amssymb}
\usepackage{epsfig}
\usepackage{rotating}
\usepackage{natbib}
\usepackage{tabularx}
\usepackage{caption}
\usepackage{soul,color}
\makeatletter
\def\hlinewd#1{%
  \noalign{\ifnum0=`}\fi\hrule \@height #1 \futurelet
   \reserved@a\@xhline}
\makeatother

\def\b#1{\mbox{\boldmath $#1$}}
\def\bl#1{\mbox{\scriptsize{\boldmath $#1$}}}

\newcommand{\logit}{{\rm logit}}

\def\cg#1{\mbox{${\cal #1}$}}
\newcommand{\tr}{^{\prime}}

\newcommand{\be}{\beta}
\newcommand{\ga}{\gamma}
\newcommand{\de}{\delta}

\newcommand{\la}{\lambda}
\renewcommand{\th}{\theta}

\begin{document}

\title{Multidimensional latent Markov models in a developmental study
  of inhibitory control and \\attentional flexibility in early childhood}

\author{Francesco Bartolucci\footnote{Department of Economics, Finance and Statistics,
University of Perugia, Via A. Pascoli 20, 06123 Perugia, Italy.}
\hspace*{0.05cm} and Ivonne L. Solis-Trapala\footnote{Division of
Medicine, Faraday Building, Lancaster University, Lancaster, LA1
4YB, UK.}} \maketitle\vspace*{-0.75cm}

\begin{abstract}
We demonstrate the use of a multidimensional extension of the latent
Markov model to analyse data from studies with correlated binary
responses in developmental psychology. In particular, we consider an
experiment based on a battery of tests which was administered to
pre-school children, at three time periods, in order to measure
their inhibitory control and attentional flexibility abilities. Our
model represents these abilities by two latent traits which are
associated to each state of a latent Markov chain. The conditional
distribution of the tests outcomes given the latent process depends
on these abilities through a multidimensional two-parameter logistic
parameterisation. We outline an EM algorithm to conduct likelihood
inference on the model parameters; we also focus on likelihood ratio
testing of hypotheses on the dimensionality of the model and on the
transition matrices of the latent process. Through the approach
based on the proposed model, we find evidence that supports that
inhibitory control and attentional flexibility can be conceptualised
as distinct constructs.  Furthermore, we outline developmental
aspects of participants' performance on these abilities based on
inspection of the estimated transition matrices.\vspace*{0.5cm}

\end{abstract}

\section{Introduction}\label{sec1}
A fundamental scientific challenge in neuropsychology and
developmental psychology is the search for evidence that two or more
postulated mechanisms, thought to underlie the performance of
participants on a set of tests, are separable. For example,
\cite{donohoe06} and \cite{kimberg00}  provide an instance where two
studies yield contradicting conclusions regarding the separability
of the psychological constructs inhibitory control and working
memory.  Whilst \citeauthor{donohoe06}~'s findings suggest that
inhibitory control is a construct that overlaps with, but is
separate to working memory, \citeauthor{kimberg00} argue against the
theory that there is an inhibitory mechanism above and beyond
working memory.

The characterisation of the latent trait underlying the performance of
participants on a set of tests in terms of a single unidimensional
latent variable would imply that all items measure a common construct.
This model can be contested by specifying a multivariate latent trait
which would postulate that the mechanisms underlying performance are
distinct theoretical concepts.

Key changes in cognitive development take place during childhood. In
this paper we assess the separability of the executive functions
inhibitory control and attentional flexibility in the context of a
study involving young children. Longitudinal studies that aim to study
developmental aspects of cognition typically involve the
administration of several tasks, to participants, where each task is
presented in blocks of trials during a single session. This session is
subsequently replicated over fixed periods of time, for example every
6 months, when it is believed that key changes in child cognition
might have occurred. A common protocol approach involves the recording
of participants' success or failure on every single trial performed
during the length of the study. This yields a sequence of
correlated binary responses for each participant.

In the presence of dichotomously-scored items, the problem above may
be translated into that of assessing the dimensionality of an Item
Response Theory (IRT) model, such as the Rasch model
\citep{rasch61}, also known as the one-parameter (1PL) logistic
model, or the two-parameter logistic (2PL) model \citep{birn68}.
This problem has been debated since long time in the statistical and
psychometric literature. Relevant contributions in this sense are
those of \cite{martin:1973}, \cite{van:simp:1982, van:two:1982},
\cite{glas:1995}, \cite{Verh:2001},
\cite{Chri:Bjor:Krei:Pete:test:2002} and, more recently,
\cite{bart:07}. In the latter paper, in particular, a class of
multidimensional IRT model is proposed which is based on a 2PL
parameterisation of the probability of responding correctly to each
item and on the assumption that the latent traits follow a
multivariate discrete distribution with an arbitrary number of
support points.  \cite{bart:07} showed how to test for
undimensionality by a Wald statistic, or equivalently a likelihood ratio
(LR) statistic between a bidimensional and a unidimensional model, which are
formulated on the basis of the above assumptions. He also showed how
this procedure may be extended to assess the number of distinct latent
traits measured by the test items and, consequently, to cluster these
items in homogeneous groups.

The above approaches, and in particular that of \cite{bart:07}, can be
directly applied to assess the dimensionality of the psychological
response process described above, provided that responses are
collected at a single occasion and that the ability of each subject in
responding correctly to each item remain constant during the
test. This excludes the cases in which the same set of items is
repeatedly administered to the same subjects during a single testing
session, which is subsequently replicated at a certain number of
occasions separated by a suitable interval of time. This scheme is
very common in psychological applications as the one which motivates
this paper. The particular feature of this scheme is that a subject
may evolve in his/her latent characteristics between occasions and
this is not taken into account in the multidimensional IRT models
mentioned above. This evolution in time is typically due to tiring
effects, learning-through-training and other developmental phenomena.

In this paper, we propose a multidimensional model for the analysis of
data deriving from design protocols that use a repeated measurement
scheme at each occasion of a longitudinal study. The basic tool is the
latent Markov (LM) model of \cite{wig73}, which may be seen as an
extension for longitudinal data of the latent class model
\cite{laz-hen68} in which each subject is allowed to move between
latent classes during the period of observation. For a detailed
description of the LM model see \cite{lan-pol02} and \cite{bar06}; for
a description with an IRT perspective see \cite{bart:08}. In our
formulation: (i) the response variables are conditionally independent
given a sequence of latent variables that follows a first-order Markov
chain which is partially homogeneous; (ii) each state of the chain is
associated to a certain level of each latent trait measured by the
test items; (iii) the distribution of each response variable depends
on these latent states through the same 2PL parameterisation adopted
by \cite{bart:07}. Evolution of the subjects with respect to these
latent traits depends on the transition matrices of the model
and, on the basis of these matrices, we can test the hypothesis
that the subjects always remain in the same state. Obviously, by
exploiting the proposed model, we can also perform an LR test for the
hypothesis of unidimensionality against that of
multidimensionality. In contrast to more standard approaches, this
test takes into account the longitudinal structure of the data. We
also allow for the inclusion of covariates affecting the probability
to belong to each latent state by a parameterisation similar to that
of \cite{ver-al99}.

For the maximum likelihood (ML) estimation of the LM model, we
illustrate an EM algorithm \citep{demp-al77} which closely recalls
that illustrated by \cite{bar06}. At the E-step, the algorithm
exploits certain recursions developed within the hidden Markov
literature \citep{mac-zuc97}. We also deal with the asymptotic
distribution of the LR statistic for testing hypotheses on the
transition probabilities. Note that, under hypotheses of this type,
the parameters may be on the boundary of the parameter space and,
therefore, an inferential problem under non-standard conditions
arises \citep{sel-lia87,sil-sen04}. However, as shown by
\cite{bar06}, this LR test statistic has an asymptotic
chi-bar-squared distribution under the null hypothesis.

The remainder of this paper is organised as follows. In the next
Section we describe the psychological study that motivates this
paper. In Section \ref{sec3} we illustrate the main assumptions of
the proposed multidimensional LM model and in Section \ref{sec4} we
illustrate likelihood inference on this model, including testing
hypotheses on the dimension of the model and on the transition
matrices of the latent process. Results of the application of our
modelling approach to the study on executive function are presented
in Section \ref{sec5}. We close with a final discussion in Section
\ref{sec6}.
\section{A longitudinal study on executive function}\label{sec2}
The study conducted by \cite{shi04} aims to investigate theoretical
questions regarding the relationships between the components of the
cognitive construct {\it executive function} in young children.  The
study comprises data collected from the administration of a battery
of tests to 115 children during a {\it single} testing session. This
session was subsequently replicated {\it over} two 6-month periods
when it was believed that key changes in child cognition might have
occurred. At the first period of testing the participants' age range
was 34-55 months. In this paper we restrict our attention to two
components of executive function, namely {\it inhibitory control}
and {\it attentional flexibility}. These are two abstract concepts
that define two closely related psychological constructs. We address
two methodological issues: (i) we investigate the nature of the
interrelationship between the two constructs and (ii) we assess
developmental trends in task performance and scalability of the
tasks at various ages.

\begin{figure}[ht]
\centerline %
\makebox{\includegraphics[width=16cm]{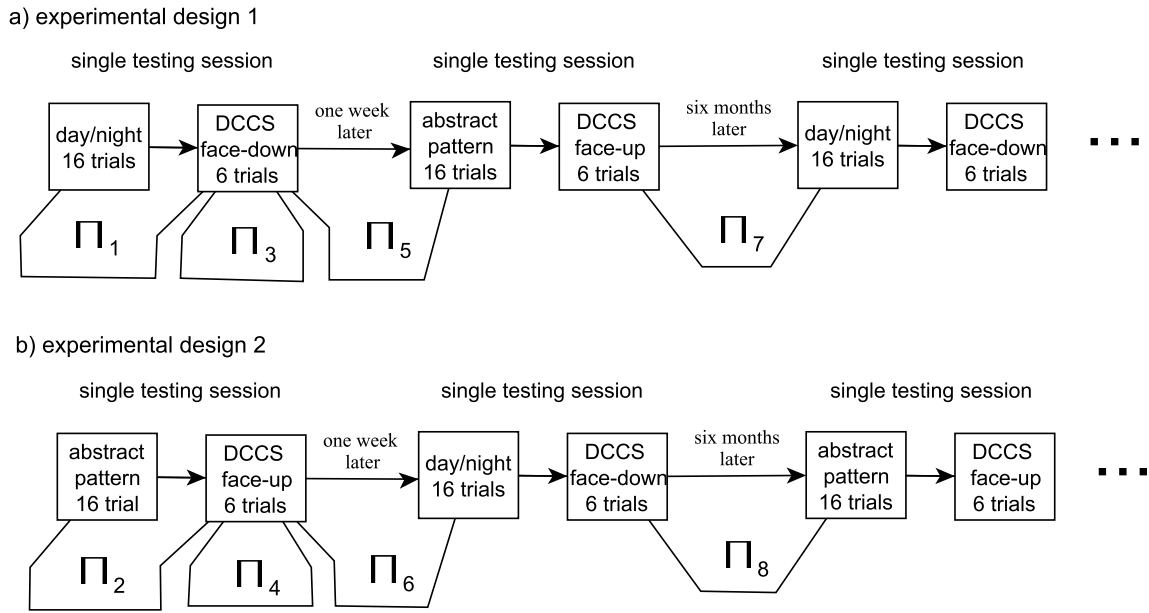}} \caption{\em
Experimental stages of participants'
      performance. The matrices $\b\Pi_r$, $r=1,\ldots,8$, contain the transition probabilities
      which are estimated within the proposed approach.}\label{fig:transmat}
\end{figure}

The response data collected for each participant consist of a sequence
of correlated binary outcomes with each component indicating a success
or failure for each trial on {\it four tasks} administered at each of
three time periods. Inhibitory control was measured by the {\it
  day/night} and the {\it abstract pattern} tasks developed by
\cite{gerstadt94}; each of these tasks was administered in blocks of
16 trials. Similarly, attentional flexibility was measured by two
versions of the {\it Dimensional change card-sort} (DCCS) tasks: the
DCCS face-up and the DCCS face-down tests \citep{zelazoetal96}. Each
of these tests was administered in blocks of 6 trials. Participants
were randomly allocated to one of two testing orderings. Half of the
group performed the tasks in the following order: day/night, DCCS
face-down, abstract pattern, DCCS face-up, whereas the other half of
the group followed the order: abstract pattern, DCCS face-up,
day/night, DCCS face-down. Thus the length of the response sequence
for each participant is 44 (16+6+16+6) at each time period. The
maximum number of observations of each subject is then $3\times 44 =
132$. For an illustration of this experiment see Figure
\ref{fig:transmat}.\vspace*{0.75cm}

\vspace*{0.75cm}

Note that the abstract pattern task is expected by design to demand
less inhibitory control skill than the day/night task. Similarly,
the face-up version of the DCCS task has lower working memory
demands than the face-down version. The rationale of the
experimental design was to administer the harder tasks previous to
their easier versions to half of the participants, and the reverse
order to the other half in order to control for potential testing
ordering effects.
\section{The multidimensional latent Markov model}\label{sec3}
Let $n$ denote the sample size, let $T_i$, $i=1,\ldots,n$, denote
the number of tasks administered to subject $i$ and let $\b
Y_i=(Y_{i1},\ldots,Y_{iT_i})$ denote the corresponding vector of
response variables, with $Y_{it}$ being a random variable equal to 1 if
subject $i$ performs correctly the task administered at occasion $t$
and 0 otherwise. From the description in the previous section it is
clear that the subjects in the sample are administered the tasks in
a different order, depending on the experimental design which is
randomly adopted for each of them. To take this into account we
introduce the index $j_{it}$, $i=1,\ldots,n$, $t=1,\ldots,T_i$, for
the type of task administered to subject $i$ at occasion $t$. In
practice, $j_{it}\in\cg J=\{1,\ldots,J\}$ for every $i$ and $t$,
with $J$ denoting the number of different types of task. In our
application $J=4$, with $j_{it}$ equal to 1 for day/night task, 2
for abstract pattern, 3 for DCCS face-down and 4 for DCCS face-up.
\subsection{Basic assumptions}
The proposed model is based on the assumption that the tasks under
consideration measure $s$ different types of latent traits or, more
specifically, abilities; then, we denote by $\cg J_d$,
$d=1,\ldots,s$, the subsets of $\cg J$ containing indices of the
tasks measuring ability of type $d$. In our application, for
instance, it is reasonable to set $s=2$, with $\cg J_1=\{1,2\}$ and
$\cg J_2=\{3,4\}$, so that day/night and abstract pattern tasks
measure the ability of type 1 and DCCS tasks measures the ability of
type 2 in both  face-down and face-up versions. Consequently to each
subject $i$ is associated an $s$-dimensional vector of abilities
which is allowed to depend on time and is denoted by $\b\th_{it} =
(\th_{it1},\ldots,\th_{its})\tr$. We also denote by
$\la_j(\b\th_{it})$ the probability that this subject performs
correctly a task of type $j$ at occasion $t$.

Following \cite{bart:07}, we adopt a 2PL parameterisation
\citep{birn68} for this conditional probability, so that
\begin{equation}
\logit[\la_j(\b\th_{it})]=\ga_j(\sum_d\de_{jd}\th_{itd}-\be_j),\quad
j=1,\ldots,J.\label{eq:multi_2pl}
\end{equation}
where $\ga_j$ is the discriminant index for the $j$th item and
$\de_{jd}$ is a dummy variable equal to 1 if $j\in\cg J_d$ and to 0
otherwise. With this notation, and taking into account that each
subject faces a specific sequence of tasks, the probability that
subject $i$ responds $y_{it}$ to the $t$-th task administered is
\begin{equation}
p(y_{it}|\b\th_{it})=\la_{j_{it}}(\b\th_{it})^{y_{it}}[1-\la_{j_{it}}(\b\th_{it})]^{1-y_{it}}.
\label{pcond}
\end{equation}
Obviously,
the simpler one-parameter (1PL) logistic parameterisation
\citep{rasch61} could also be used. In this case, we have
\begin{equation}
\logit[\la_j(\b\th_{it})]=\sum_d\de_{jd}\th_{itd}-\be_j,\quad
j=1,\ldots,J.\label{eq:multi_1pl}
\end{equation}
This parameterisation is less flexible than the above one since it
assumes that all taks have the same discriminant power, i.e.
$\ga_j=1$, $j=1,\ldots,J$. On the other hand, the resulting model is
simpler to estimate.

In order to model the dynamics of each ability across $t$, we
assume that for each subject $i$ the latent process
$\b\th_{i1},\ldots,\b\th_{iT_i}$ follows a first-order Markov chain
with state space $\{\b\xi_1,\ldots,\b\xi_k\}$, where $k$ is the
number of latent states, with initial probabilities $\pi_{ic} =
p(\b\th_{i1}=\b\xi_c)$, $c=1,\ldots,k$, and transition probabilities
$\pi_{icd}^{(t)}=p(\b\th_{it}=\b\xi_d|\b\th_{i,t-1}=\b\xi_c)$,
$c,d=1,\ldots,k$. Since we consider the abilities as the only
explanatory variables of the probability of performing correctly an
item, as in the latent Markov model of \cite{wig73} we assume that
the response variables in $\b Y_i$ are conditional independent given
$\b\th_{i1},\ldots,\b\th_{iT_i}$. The resulting
model is then an extension of the latent Markov Rasch model
described by \cite{bart:08} based on a multidimensional 2PL
parameterisation for the conditional distribution of the response
variables given the latent process and allowing for different
sequences of items between subjects. Therefore, experimental design aspects
  such as testing order can also be taken into consideration in the
  statistical analysis of the data. Moreover, by assuming the
existence of a latent Markov chain, we allow a subject to move
between latent classes in a way that depends on the transition
probabilities and hence we also generalise the multidimensional IRT
model of \cite{bart:07} which implicitly assumes that every subject
maintains the same level of each ability across time occasions.

A final point concerns how to model in a parsimonious way the initial
and transition probabilities of the latent process taking into
account the difference between subjects in terms of individual covariates
and considering the experimental design. For this aim, and
similarly to \cite{ver-al99}, we adopt the following logistic
parameterisation for the initial probabilities:
\begin{equation}
\log\frac{\pi_{ic}}{\pi_{i1}}=\b x_i\tr\b\phi_c,\qquad
c=2,\ldots,k,\quad i = 1,\ldots,n,\label{eq:par_logit}
\end{equation}
where $\b x_i$ is a vector of
covariates measured at the first
occasion. In our application, we allow these initial probabilities
to depend on the age of the subject at beginning of the study,
indicated by $a_i$ for subject $i$, and then we let $\b
x_i=(1,a_i)\tr$.

For what concerns the transition probabilities of the latent process, we specify
those transition matrices that encompass the different
transitional stages of substantive interest. For instance, for our
application, we use a first type of transition matrix within the
sequence of day/night tasks and a second type within the second of
abstract pattern tasks and so on; see Section \ref{sec6} for a
detailed description. Note that we could also allow the transition
probabilities to depend on the individual covariates through a parameterisation of
the probabilities $\pi^{(t)}_{icd}$ similar to that in
(\ref{eq:par_logit}). This however would complicate the approach and
we do not pursue this further here.

Submodels may also be considered. The most interesting is obviously
that of unidimensionality that may be formulated by requiring that
$\b\xi_c=\xi_c\b 1_k$, with $\b 1_k$ denoting a column vector of $k$
ones, so that the elements of $\b\xi_c$ are equal to each other. The
hypothesis that there is no transition between latent states at a
certain occasion is also of interest. This may be formulated by
requiring that the corresponding matrix of transition probabilities is
diagonal; see \cite{bar06} for a detailed description of these
constraints. Finally, the assumption that there is no effect of the
covariates on the initial probabilities is also of interest. In our
application this assumption may be formulated by letting $\b x_i=1$,
$i=1,\ldots,n$.
\subsection{Manifest distribution of the response variables}
In order to derive the manifest distribution of the response
variables $Y_{it}$, we consider the vector $\b C_i =
(C_{i1},\ldots,C_{iT_i})$, with $C_{it}$ being the state to which
subject $i$ belongs at occasion $t$, with $i=1,\ldots,n$,
$t=1,\ldots,T_i$. In practice, $C_{it}=c$, for $c=1,\ldots,k$, if and only
if $\b\th_{it}=\b x_c$.

The assumption of conditional independence between the elements of
$\b Y_i$ given the latent process implies that
\[
p(\b y_i|\b c_i) = \prod_j \la_{itc_{it}}^{y_{it}}(1-\la_{itc_{it}})^{1-y_{it}},
\]
where $\b y_i$ denotes a realisation of $\b Y_i$, $\b c_i$ denotes a
realisation of $\b C_i$ whose elements are indicated by $c_{it}$ and
\begin{equation}
\la_{itc}=p(Y_{it}=y_{it}|\b\th_{it}=\b\xi_c),\label{prob_la}
\end{equation}
which is defined on the basis of (\ref{pcond}).
Consequently, the {\em manifest
distribution} of $\b Y_i$ may be expressed as
\begin{equation}
p(\b y_i) = \sum_{\bl c_i} p(\b y_i|\b c_i)p(\b c_i),\label{eq:sum_prob}
\end{equation}
where the sum is extended to all the possible $k^{T_i}$
configurations of $\b C_i$. Clearly, the probability $p(\b c_i)$
depends on the initial and transition probabilities $\pi_{ic}$
and $\pi_{icd}^{(t)}$.

Avoiding the sum in (\ref{eq:sum_prob}), the manifest probability
$p(\b y_i)$ may be efficiently computed by exploiting certain
recursions known in the hidden Markov literature \citep{mac-zuc97}
and which may be efficiently implemented by using matrix notation;
we refer to \cite{bar-al07} for details. Let $\b\pi_i$ denote the
initial probability vector with element $\pi_{ic}$, $c=1,\ldots,k$,
and let $\b\Pi_{it}^{(t)}$ denote the transition probability matrix
with elements $c,d=1,\ldots,k$, $c\neq d$. For a fixed vector $\b
y_i$, also let $\b l_{it}$ denote the column vector with elements
$p(\b\th_{it}=\b\xi_c,Y_{i1}=y_{i1},\ldots,Y_{it}=y_{it})$ for
$c=1,\ldots,k$. For $t=1,\ldots,T_i$, this vector may be computed by
using the recursion
\begin{equation}
\b l_{it} = \left\{\begin{array}{ll}
\textrm{diag}(\b\la_{i1}) \b\pi_i & \mbox{ if } j=1,\\
\textrm{diag}(\b\la_{it})[\b\Pi^{(t)}]\tr \b l_{i,t-1} & \mbox{ otherwise},\\
\end{array}\right. \label{rec}
\end{equation}
where $\b\la_{it}$ denotes the column vector with elements $\la_{itc}$, $c=1,\ldots,k$. Once (\ref{rec}) has
been computed for $t=1,\ldots,T_i$, we obtain $p(\b x)$
as the sum of the elements of $\b l_{iT_i}$.

\section{Likelihood inference on the model parameters}\label{sec4}
Let $\b\eta$ denote the vector of all model parameters, i.e. the
item parameters $\be_j$ and $\ga_j$, the support points $\b\xi_c$ of the latent distribution,
the parameters $\b\phi_c$ for the initial probabilities of the latent
process and the transition probabilities $\pi_{icd}^{(t)}$.

As usual, we assume that the subjects in the sample have response
patterns independent from one another, so that the {\em log-likelihood} of
the LMR model may be expressed as
\[
\ell(\b\eta) = \sum_i\log[p(\b y_i)],
\]
where $p(\b y_i)$ is computed as a function of $\b\eta$ by using
recursion (\ref{rec}). We can maximise $\ell(\b\eta)$ by means of
the EM algorithm \citep{demp-al77} implemented as in \cite{bar-al07}
and which we briefly illustrate in the following.

\subsection{Log-likelihood maximisation}\label{sec4.1}
First of all, we need to consider the so-called {\em complete data
log-likelihood} that may be expressed as
\begin{equation}
\begin{array}{ll}
\ell^*(\b\eta)&=\sum_i[\sum_cw_{i1c}\log(\pi_{ic})+ \sum_{t=2}^{T_i}\sum_c\sum_d
z_{icd}\log(\pi_{icd}^{(t)}) + \cr\vspace*{-0.2cm}\cr
&\hspace*{1.1cm}+\sum_{t=1}^{T_i}\sum_cw_{itc}\{y_{it}\log(\la_{itc})+(1-y_{it})\log(1-\la_{itc})\}],
\end{array}\label{comp-lk}
\end{equation}
where $w_{itc}$ is a dummy variable equal to 1 if subject $i$ is in latent state $c$
at occasion $t$ (i.e. $C_{it}=c$), $z_{icd}^{(t)}=w_{i,t-1,c}w_{itd}$ is a dummy variable
equal to 1 if subject $i$ moves from the $c$-th to the $d$-th latent state
at occasion $t$ (i.e. $C_{i,t-1}=c$ and i.e. $C_{it}=d$) and the probabilities in $\la_{itc}$ are defined in (\ref{prob_la}).

Obviously, the dummy variables in $\ell^*(\b\eta)$ are not known and then this log-likelihood
is exploited within the EM algorithm to maximize the incomplete data
log-likelihood $\ell(\b\eta)$ by alternating the following two steps
until convergence:
\begin{itemize}
\item {\em E-step}: compute the conditional expected value of the dummy variables $w_{itc}$ and $z_{icd}$
given the observed data $y_{it}$, the covariates $\b x_{it}$ and the current
estimate of $\b\eta$;
\item {\em M-step}: update the
estimate of $\b\eta$ by maximizing the expected value of the complete log-likelihood
$\ell^*(\b\eta)$; this is defined as in (\ref{comp-lk}) with each dummy variable involved
in this expression substituted by the corresponding expected value
computed at the E-step.
\end{itemize}

The E-step may be performed by using certain recursions similar to
(\ref{rec}). In order to udpate the parameters $\be_j$, $\ga_j$ and
$\b\xi_c$, the M-step requires to run an algorithm to maximize the
weighted likelihood of a logistic model, which is easily available.
A similar algorithm is necessary to update the parameters $\b\phi_c$
for the initial probabilities, whereas an explicit expression is
available to update the transition probabilities $\pi_{icd}^{(t)}$.
For a detailed description on how to implement these steps, see
\cite{bar06}, \cite{bar-al07} and \cite{bart:08}.

The algorithm requires to be initialized by choosing suitable
starting values for the parameters in $\b\eta$. Trying different
sets of starting values is useful to detect multimodality of the
likelihood, which often arises for latent variable models. These
starting values may be generated by a random rule and then we take,
as ML estimate of the parameters, the value of $\b\eta$ which at
convergence gives the highest value of $\ell(\b\eta)$. This estimate
is denoted by $\hat{\b\eta}$.
\subsection{Model seleciton and testing hypotheses on the
parameters}\label{sec4.2}
Obviously, analysing a dataset through the model described above
requires to choose the number of latent states $k$. Following the
current literature on latent variable models, and the related
literature on finite mixture models \cite[see][]{mcla:peel:00}, we
suggest to rely on the Bayesian Information Criterion (BIC) which
was proposed by \cite{schw78}. According to this criterion, we
choose the value of $k$ corresponding to the minimum of the index
\begin{equation}
BIC_k = -2\hat{\ell}_k+g_k\log(n),\label{bic}
\end{equation}
where $\hat{\ell}_k$ denotes the maximised log-likelihood and $g_k$
is the corresponding number of parameters.

Note that the penalization term in (\ref{bic}) only takes the sample
size into account. In order to take into account the overall number
of observations, corresponding to $\sum_iT_i$, we can alternatively
use the criterion BIC$^*$ based on the minimisation of the index
\[
BIC_k^* = -2\hat{\ell}_k+g_k\log(\sum_iT_i).
\]

Once the number of latent states has been chosen, a hypothesis $H_0$
on the parameters $\b\eta$ may be tested by the likelihood ratio
(LR) statistic
\[
D = -2[\ell(\hat{\b\eta}_0)-\ell(\hat{\b\eta})],
\]
where $\hat{\b\eta}_0$ is the ML estimate of $\b\eta$ under $H_0$;
this estimate may also be computed by using the EM algorithm
illustrated above.

When standard regularity conditions hold, the asymptotic null
distribution of the statistic $D$ is a chi-squared distribution with
a number of degrees of freedom equal to the number of independent
constraints used to express $H_0$. In particular, through this
procedure we may test hypotheses on the transition probabilities of
the latent process. As already mentioned, standard asymptotic
results on the distribution of LR test statistic are not ensured to
hold in this case. This happens, for instance, for the hypothesis
that a transition matrix is diagonal and thus transitions between
latent states is not possible. However, by using certain results on
constrained statistical inference \citep{sel-lia87,sil-sen04}, it
was shown by \cite{bar06} that the null asymptotic distribution of
$D$ is of chi-bar-squared type \citep{shap88,sil-sen04}. This is a
mixture of standard chi-squared distribution with weights that may
be computed by a simple rule.
\section{Results}\label{sec5}
In this section we investigate models that describe the
developmental and dimensional aspects of the data collected to
measure inhibitory control and attentional flexibility,
described in Section \ref{sec2}. The tasks that measure the above
abilities correspond to $J=4$ types of item, namely: (i) day/night, (ii) abstract pattern, (iii) DCCS
face-down and (iv) DCCS face-up. As we described in detail in Section \ref{sec3}, our modelling approach
relies on a latent Markov process. The parameters of the distribution of this process
are specified as follows:
\begin{itemize}
\item[(i)] the initial probabilities of the latent process depend
on age according to (\ref{eq:par_logit});
\item[(ii)] there are 8 different transition matrices of the latent
  process that model participants' ability transition during the
  period of experimentation. The experimental stages of
    substantive interest and their corresponding transition
    probability matrices are depicted in
  Figure~\ref{fig:transmat}. Each testing session in the
    study involves two transition matrices. The first matrix measures
    changes in performance within the inhibitory control sequence of
    trials and transition to the attentional flexibility task, whereas the
    second one measures transitions within the attentional
    flexibility sequence of trials. Thus, changes in performance
    during the two types of testing sessions yield four transition matrices, $\b\Pi_1$
    and $\b\Pi_3$ for the experimental design where harder tasks precede
    easier tasks, and $\b\Pi_2$ and $\b\Pi_4$ for the alternative
    experimental design. Transition matrices $\b\Pi_5$ and $\b\Pi_6$
    represent changes in abilities between sequences over a week
    period of time, and similarly, matrices $\b\Pi_7$ and $\b\Pi_8$ model
    changes over six-month time periods for the two experimental
    designs.
\end{itemize}

To complete our model specification we initially assume that the
probability of responding correctly to each item is unconstrained,
so that we have a specific parameter $\la_{jc}$ for the probability
of responding correctly to each item of type $j$, $j=1,\ldots,J$, given each
possible latent state $c$, $c=1,\ldots,k$. Subsequently we test
whether the 2PL parameterisation based on (\ref{eq:multi_2pl}) is
reasonable.

We next turn to the problem of choosing an LM model that provides an
appropriate representation of the dependence structure. We first
address the choice of the number of latent classes. We consider four
models with 1 to 4 latent classes and select the fitted model with
smallest values of the Bayesian information criteria BIC and BIC*
where, as shown earlier, BIC is penalised by a term equal to the
sample size 115 and BIC* is penalised by the total number of single
trials equal to 12798. Table~\ref{tab:latclass} displays the values
of the log-likelihood function evaluated at the maximum likelihood
estimate, the BIC and BIC* indices for model comparison.

\begin{table}[ht]\centering
\caption{{\it  Comparison of models with various
values of $k$ by maximum log-likelihood, BIC and BIC* indices}}\label{tab:latclass}
\begin{tabular}{rrrrr}\hlinewd{1.3pt}\\
$k$   &\multicolumn1c{log-likelihood}& \multicolumn1c{no. of par.} &    \multicolumn1c{BIC}&    \multicolumn1c{BIC$^*$}\\\hline
1   &-7050.1&   4&  14119.0 &  14138.0\\
2   &-3907.5&       26& 7938.4& 8060.9\\
3   &-3545.3&   64& 7394.3& 7695.9\\
4   &-3461.1&   118&    7482.1& 8038.1\\
\hline
\end{tabular}
\hspace{.2cm}
\begin{minipage}{16.5cm}{
{\footnotesize Note: No. of par. is the number of
  parameters; BIC=Bayesian Information Criterion with penalization
  term equal to the sample size 115; BIC*=BIC with extended
  penalization term to the total number of single trials 12798.}}
\end{minipage}\vspace*{0.75cm}
\end{table}

On the basis of the results in Table~\ref{tab:latclass}, we
select a model with $k=3$ latent states. This model may be
simplified by testing hypotheses on the equality between some of
the transition matrices deriving from the experimental stages
depicted in Figure~\ref{fig:transmat}. For instance, it is of
interest to test whether patterns of participants' performance
within each of the two tasks of inhibitory control are comparable.
Formally, this involves testing the hypothesis of equality between
the transition matrices $\b\Pi_1$ and $\b\Pi_2$. Similarly, comparison between
matrices $\b\Pi_5$ and $\b\Pi_6$ and matrices $\b\Pi_7$ and $\b\Pi_8$ is important because
rejection of the hypothesis of equality of these matrices would
suggest the presence of task order effects.

Table~\ref{tab:tranhyp} summarises the results of testing the above
and other relevant hypotheses on the transition matrices. We
highlight in boldface those constrained models for which the BIC and
BIC$^*$ indices are minimum among fitted models with comparable number
of parameters.

\begin{table}[ht!]\centering\vspace*{.75cm}\caption{{\em Hypotheses testing on the transition
      matrices that represent changes in participants' performance
      during the experimental stages depicted in Figure~\ref{fig:transmat}}\label{tab:tranhyp}}
\begin{tabular}{lrrrr}\hlinewd{1.3pt}\\
Hypothesis &    log-likelihood & No. par. &    BIC&    BIC*\\\hline
$\b\Pi_1=\b\Pi_2$&     -3556.3&    58& 7387.9& 7661.2\\
$\b\Pi_3=\b\Pi_4$&      -3557.2&    58& 7389.6& 7662.9\\
$\b\Pi_5=\b\Pi_6$&    {\bf -3555.2}&  {\bf 58}&
{\bf 7385.6}&   {\bf 7658.9}\\
$\b\Pi_7=\b\Pi_8$&      -3555   &58&    7385.3&
7658.6\\\hline
$\b\Pi_1=\b\Pi_2$, $\b\Pi_5=\b\Pi_6$&    -3569.8&    52& 7386.3& 7631.3\\
$\b\Pi_3=\b\Pi_4$, $\b\Pi_5=\b\Pi_6$&    -3562.2&    52& 7371.1& 7616.1\\
$\b\Pi_5=\b\Pi_6$, $\b\Pi_7=\b\Pi_8$&  {\bf -3554.8}&  {\bf
  52}&  {\bf 7356.4}&   {\bf 7601.4}\\\hline
$\b\Pi_1=\b\Pi_2$, $\b\Pi_5=\b\Pi_6$, $\b\Pi_7=\b\Pi_8$& -3573.5&    46& 7365.2& 7582\\
\mathversion{bold}$\Pi_3=\Pi_4$, $\Pi_5=\Pi_6$, $\Pi_7=\Pi_8$\mathversion{normal}& {\bf
-3566.7}& {\bf 46}&   {\bf 7351.6}&   {\bf 7568.3}\\\hline
$\b\Pi_1=\b\Pi_2$, $\b\Pi_3=\b\Pi_4$, $\b\Pi_5=\b\Pi_6$, $\b\Pi_7=\b\Pi_8$ & {\bf
  -3572.2}& {\bf 40}&   {\bf 7334.2}&{\bf   7522.7}\\\hline
$\b\Pi_1=\b\Pi_2=\b I_3$&  -4014&  34& 8189.2& 8349.4\\
$\b\Pi_3=\b\Pi_4=\b I_3$&  -3709.9&    34& 7581.1& 7741.3\\
$\b\Pi_5=\b\Pi_6=\b I_3$&  -3641.5&    34& 7444.3& 7604.5\\
$\b\Pi_7=\b\Pi_8=\b I_3$&  -3656.6&    34&
7474.6& 7634.8\\\hline
\end{tabular}
\begin{minipage}{16.5cm}{
{\footnotesize Note: No. par.= Number of
  parameters; BIC=Bayesian Information Criterion with penalization
  term equal to the sample size 115; BIC$^*$=BIC with extended
  penalization term to the total number of single trials 12798.
Constrained models for which BIC and BIC$^*$ are minimum among fitted
models with comparable number of parameters are highlighted in
boldface; $\b I_3=3\times3$ identity matrix.}}
\end{minipage}\vspace*{0.75cm}
\end{table}

Based on the results presented in Table~\ref{tab:tranhyp}, we adopt
a constrained model where the matrices $\b\Pi_1$ and $\b\Pi_2$ are equal
and so are matrices $\b\Pi_5$ and $\b\Pi_6$, and $\b\Pi_7$ and $\b\Pi_8$. Further reduction
is not plausible as can be seen from the fact that constraining the
transition matrices to be identity matrices yields larger values of
BIC and BIC$^*$. Hence, we can represent the dependence structure
through four transition matrices. The first two transition matrices
comprise patterns of performance within the inhibitory control  and
within the attention flexibility items respectively. The third
transition matrix yields information on changes in performance after
a week's time; the fourth transition matrix describes
developmental changes that take place in 6 months' time.

Furthermore, we tested the dependence of the initial probabilities
on participant's age. We observe that the value of the BIC increases
from 7334.2 to 7446.4 when the dependence is removed. Thus we conclude
that age has a significant effect on the initial probabilities.

Another important aspect in the search of an appropriate latent
Markov model concerns the parameterisation of the conditional
probabilities of success at performing the task given the latent
state. We explore the 1PL and 2PL parameterisations,
contrasting between the unidimensional and the bidimensional cases;
see equations (\ref{eq:multi_2pl}) and
(\ref{eq:multi_1pl}). We adopt a bidimensional 2PL model which yields the smallest values of BIC compared with
its competing models, as can be seen from Table~\ref{tab:param}. The
first ability underlies the inhibitory control tasks, whereas the
second ability is the basis of the attentional flexibility tasks.
This completes our model choice.\vspace*{0.75cm}

\begin{table}[ht]\centering\caption{{\em Model comparison for unidimensional and
      bidimensional parameterisations of the Rasch and the
      2PL logistic models }}\label{tab:param}
\begin{tabular}{lrrrr}\hlinewd{1.3pt}\\
\multicolumn1c{Model}& \multicolumn1c{log-likelihood} & \multicolumn1c{No. par.} &  \multicolumn1c{BIC} & \multicolumn1c{BIC$^*$}\\\hline
Rasch unidimensional&           -3696.0 & 34 &    7553.2& 7713.4\\
Rasch bidimensional&            -3580.6 &36 &   7332.0 &  7501.6\\
2PL unidimensional&         -3679.8 &37&    7535.2& 7709.6\\
2PL bidimensional& -3572.3& 38& 7325.0 & 7504.0\\\hline
\end{tabular}
\begin{minipage}{16cm}{
{\footnotesize Note: 2PL=two-parameter logistic model;  No. par.=
Number of
  parameters; BIC=Bayesian Information Criterion with penalization
  term equal to the sample size 115; BIC$^*$=BIC with extended
  penalization term to the total number of single trials 12798. }}
\end{minipage}
\end{table} \vspace*{0.75cm}

In summary, we adopt a latent Markov model with three latent
variables. The initial probabilities of the latent process depend on
participant's age and there are four transition probability matrices
whose interpretation was discussed earlier. For this model, the
conditional probability of responding correctly to each item given
the latent process is parameterised as in a 2PL
bidimensional model \cite{bart:07}. Here, the first dimension represents inhibitory
control and the second one attentional flexibility. Under the
selected model the estimated ability parameters are summarised in
Table~\ref{tab:abilpar}.

\begin{table}[ht]\centering\vspace*{0.75cm}\caption{\it Ability parameter estimates for each latent
      state}\label{tab:abilpar}
\begin{tabular}{lrrr}\hlinewd{1.3pt}\\
Dimension& latent state 1 & latent state 2 & latent state 3\\\hline
Inhibitory control& -5.454&    0.040& 5.050\\
Attentional flexibility &0.145&    -35.145&   6.176\\\hline
\end{tabular}
\vspace*{0.75cm}
\end{table}

The latent states are ordered to reflect lowest to highest ability for
both dimensions. Note that the estimated ability parameters
corresponding to attentional flexibility are large in absolute value due to the fact
that a large number of participants tended to score either zero or one
during the full 6-item sequences. As noted by the experimenter, this
is not a particular characteristic of this data set; but other
studies \citep[e.g.][]{zelazoetal96} report a similar phenomenon.

Table~\ref{tab:itempar} displays estimates of the discriminant and
difficulty parameters associated with each item. These estimates
reveal that the abstract pattern task is considerably easier and
less discriminant than the day/night task. On the other hand, the two
DCCS tasks are comparable in level of difficultly, but the DCCS
face-up task discriminates better than the face-down version of this
task. These results confirm the experimenter's hypotheses
\citep[see][]{shi04}, namely that participants would perform better
at the abstract pattern and the DCCS face-up tasks than at their
counterparts. Furthermore, in the case of the inhibitory control tasks, the
abstract pattern task was regarded as a control for the day/night
task, in the sense that it is expected to make less inhibitory demands; therefore,
our finding that the day/night task is a better discriminator is reassuring.

\begin{table}[ht]\centering
\vspace*{0.75cm}\caption{\it Discriminant and difficulty parameter estimates
      associated with each item}\label{tab:itempar}
\begin{tabular}{lrr}\hlinewd{1.3pt}\\
     &  \multicolumn1c{discriminant} &  \multicolumn1c{difficulty}\\
\multicolumn1c{Item} &  \multicolumn1c{index} &  \multicolumn1c{level}\\\hline
day/night&  1.000&  0.000\\
abstract pattern& 0.610 & -4.880\\
DCCS face-down  &1.000 &    0.000\\
DCCS face-up &  2.744& 0.107\\\hline
\end{tabular}
\vspace*{0.75cm}
\end{table}

Based on the estimates reported in
Tables~\ref{tab:abilpar}~and~\ref{tab:itempar} we can calculate the
conditional probabilities of success to respond to an item for a
given participant given his/her latent state; see  equation (\ref{eq:multi_2pl}). These probabilities
are presented in Table~\ref{tab:probab}.

\begin{table}[ht]\centering
\vspace*{0.75cm}\caption{\em Estimated Conditional probabilities of
      responding correctly to each item given latent state}\label{tab:probab}
\begin{tabular}{lrrr}\hlinewd{1.3pt}\\
\multicolumn1c{Item} &  \multicolumn1c{latent state 1} & \multicolumn1c{latent state 2} & \multicolumn1c{latent state 3}\\\hline
day/night& 0.0043& 0.5101 & 0.9936\\
abstract pattern& 0.4132& 0.9527 &0.9977\\
DCCS face-down& 0.5362  &0.0000 &    0.9979\\
DCCS face-up&   0.5264& 0.0000 & 1.0000\\\hline
\end{tabular}
\end{table}

A noteworthy fact is that the participants who fall in the latent
state 1 have a fifty percent probability of success in the DCCS
tasks but low probabilities of success in the day/night task. The
interpretation of the ability on the DCCS tasks for such
participants must be made with care, since the experimenter has
reported that she believes that some participants adopted a
strategy, i.e.  alternating the card placement regardless of the
game's rules, to perform the card tasks that led to a medium size
number of successes in the 6-trial sequences.
Importantly, our analysis strongly supports the hypothesis that there are two
separable abilities, namely inhibitory control and attentional
flexibility, underlying performance at the items administered to
participants. We next address developmental questions regarding
these abilities by exploring the transition probabilities of the
assumed latent process.

We examine the effect of age on performance by modelling the initial
probabilities of the latent process in a logistic regression with
age as explanatory variable. The estimates of the regression
coefficients of age, corresponding to the logistic curves of the
initial probabilities for the second and third latent states, are
0.111 and 0.361 respectively. That is, the model classifies older
participants into states of higher performance. Averaged over all the
subjects in the sample, the three latent states have initial probabilities equal to
0.261, 0.360 and 0.379. The second and third state have comparable dimension,
whereas the first is smaller and contains around one fourth of the sample.

Table~\ref{tab:esttran} reports the estimated transition
probabilities matrices. Recall that earlier in this section we identified four transition
matrices to represent the change patterns on participants'
performance. The matrix denoted by ``within IC items'' in
Table~\ref{tab:esttran} encompasses changes in performance within the
inhibitory control items and transitions to the attention flexibility items. It corresponds to $\b\Pi_1=\b\Pi_2$.
The ``within AF items'' represents performance changes within the attention flexibility items only and corresponds to
$\b\Pi_3=\b\Pi_4$. The remaining
two transition matrices represent transition of states after a week
and six-month periods respectively, that is $\b\Pi_5=\b\Pi_6$ and $\b\Pi_7=\b\Pi_8$.

\begin{table}[ht]\centering\vspace*{0.75cm}\caption{\it Estimated transition probability
      matrices}\label{tab:esttran}
\begin{tabularx}{16cm}{c X c c c X c c c}\hlinewd{1.3pt}\\
&&\multicolumn3c{within IC items}&& \multicolumn3c{within AF
items}\\  \cline{3-5}\cline{7-9} latent state &&  1 &        2 &
3&   &                     1  &         2 &
3\\\cline{1-1}\cline{3-5}\cline{7-9}
  1&&    0.9809 &    0.0191&      0.0000  &   &                 0.5883 &     0.2211 &     0.1905\\
  2&&   0.0228 &      0.9146 &    0.0626 &&               0.0049&      0.9951 &   0.0000  \\
  3&&   0.0114 &     0.0372&      0.9514&&               0.0226&    0.0022  &    0.9753 \\\hline
&&\multicolumn3c{between AF and IC}&&\multicolumn3c{ between AF and IC}\\
&&\multicolumn3c{a week later}&&\multicolumn3c{6 months later}\\
\cline{3-5}\cline{7-9} latent state&&   1   &      2 &        3 &&
1 &         2  &           3\\ \cline{1-1}\cline{3-5}\cline{7-9}
1 &&       0.0000 &     0.5763 &     0.4237 &&                 0.3002&      0.6997&         0.0000 \\
2 &&     0.0609&   0.0000  &    0.9391  &&                      0.1795&       0.1165 &       0.7040\\
3 &&       0.0000  &    0.0000 &           1.0000 &&  0.1221 &
0.1362& 0.7416\\\hline
\end{tabularx}
\begin{minipage}{16cm}{
{\footnotesize Note: IC=inhibitory control; AF=attentional
flexibility.}}
\end{minipage}\vspace*{0.25cm}
\end{table}

The within IC items transition matrix show
evidence of a tiring effect as reflected by some tendency of
participants to move from higher to lower states within the
sequences of inhibitory control trials. In contrast, the within AF
items transition matrix primarily indicates the presence of learning
effects as participants tend to move from the first to the higher
states. Similarly, there is a learning effect during the transition
after a week's time. Finally, the probability matrix concerning
transitions over a six-month period shows strong developmental
effects for participants in states 1 and 2. There are also some
evidence of participants moving back from the second and third
states to lower states.

A final point concerns the computation effort involved in the
analysis based on the LM model described in this paper. We found
that, although estimation is based on the EM algorithm which may
require a large number of iterations, fitting the model on the data
described above is rather fast even with a large number of latent
states. This is because we use an efficient implementation based on
recursions taken from the hidden-Markov literature. Implementing
these recursions is rather easy in mathematical packages such as
{\sc Matlab}. On the other hand, we found that for the data
considered here, most of the fitted models have a multimodal
likelihood. It is then crucial to try different starting values for
the EM algorithm as described at the end of Section \ref{sec4.1}. We
make our {\sc Matlab} implementation of the EM algorithm, and the
related routine to choose the starting values, available to the
readers upon request.

\section{Discussion}\label{sec6}

We have developed an approach based on a multidimensional latent Markov
model suitable for the analysis of long binary response
sequences that measure two or more latent traits.

For each subject in the sample, our approach relies on Markov chains with transition probabilities
that are of substantive interest, in order to measure dynamic changes
in the responses during repeated trials and over fixed periods of
time. Hence the applicability of the method lies on studies where the
aim is at studying developmental or detrimental changes over time in
an intuitive way. Importantly, we show how to test the dimensionality of the
problem by developing valid LR tests.

The modelling approach has been illustrated through the analysis of data
from a developmental study. Throughout the statistical analysis we emphasized the
importance of taking into consideration experimental aspects of the
study, for instance to account for potential ordering effects. Our main
finding was evidence that supports that the tasks measure two
distinct psychological constructs, namely inhibitory control and
attentional flexibility. An obvious extension of the applicability of
our method would be to investigate hypotheses on the relationship
between other executive skills such as
working memory and planning.

In addition, we were able to demonstrate developmental aspects of the
two abilities. In particular that these abilities develop at an early
age, as the covariate age played an important role at performance. The
dynamics within sequences were different among the two types of tasks
considered. The presence of tiring effects was clear within the inhibitory control
sequences. In contrast, participants where either able or unable to
perform the attentional flexibility task. It is plausible that a third
group showed a spurious medium probability of success at this
task. Further investigation on the nature of this task is warranted.

A limitation of our approach is the assumption of a first order Markov
chain for the latent process. However, this assumption seems to be suitable
for the data analysed in this article. Indeed, a quick inspection of time
dependence through the assumption of a continuous latent process
favours a short-range dependence. Note also, that we do not require to
make any distributional assumptions on the latent process as opposed
to the conventional Gaussian assumption made for continuous processes.
\bibliography{biblio}
\bibliographystyle{apalike}
\end{document}